# Pulmonary Fissure Segmentation in CT Images Based on ODoS Filter and Shape Features

Yuanyuan Peng,[1,2✉] Pengpeng Luan,[1] Hongbin Tu,[1] Xiong Li,[3] Ping Zhou,[4]

[1] School of Electrical and Automation Engineering, East China Jiaotong University, Nanchang 330000, People's Republic of China
[2] School of Computer Science, Northwestern Polytechnical University, Xi'An 710000, People's Republic of China
[3] School of Software, East China Jiaotong University, NanChang 330000, People's Republic of China
[4] College of Biology, Hunan University, Changsha 410000, People's Republic of China

**Correspondence**

School of Electrical and Automation Engineering, East China Jiaotong University, Nanchang 330000, People's Republic of China
✉Email:3066@ecjtu.edu.cn

## Abstract

Priori knowledge of pulmonary anatomy plays a vital role in diagnosis of lung diseases. In CT images, pulmonary fissure segmentation is a formidable mission due to various of factors. To address the challenge, an useful approach based on ODoS filter and shape features is presented for pulmonary fissure segmentation. Here, we adopt an ODoS filter by merging the orientation information and magnitude information to highlight structure features for fissure enhancement, which can effectively distinguish between pulmonary fissures and clutters. Motivated by the fact that pulmonary fissures appear as linear structures in 2D space and planar structures in 3D space in orientation field, an orientation curvature criterion and an orientation partition scheme are fused to separate fissure patches and other structures in different orientation partition, which can suppress parts of clutters. Considering the shape difference between pulmonary fissures and tubular structures in magnitude field, a shape measure approach and a 3D skeletonization model are combined to segment pulmonary fissures for clutters removal. When applying our scheme to 55 chest CT scans which acquired from a publicly available LOLA11 datasets, the median $F_1$-score, False Discovery Rate (*FDR*), and False Negative Rate (*FNR*) respectively are 0.896, 0.109, and 0.100, which indicates that the presented method has a satisfactory pulmonary fissure segmentation performance.

## 1. Introduction

Pulmonary fissures are composed of a double layer of visceral pleura that separate human lungs into five lobes [1]. On the basis of anatomy, the left hung is separated into two parts by an oblique fissure, whereas the right lung is separated into three parts by a horizontal fissure and an oblique fissure [2]. For the clinical diagnosis, pulmonary fissure completeness is useful in lung disease assessment [3,4,5,6]. However, pulmonary fissure segmentation is a formidable mission due to various of factors such as intensity variability, partial volume effect and pathological deformation [7,8].

Numerous studies have been presented for pulmonary fissure segmentation, which can be composed of three different categories. The first category mainly employed the shape and structure information from fissure itself, such as ridgeness measure model [9], line-enhancing operator [10], multiple section model [11], DoS filer [12], ODoS filter [13], Hessian-based filter [14], and directional derivative of plate filter [15]. These fissure-based methods have great difficulties in handing deformed and disrupted fissures, which are caused by complex lung diseases, noise and pathological deformation. The second strategy generally exploited lung anatomical knowledge to identify pulmonary fissures under different frameworks [16], such as alpha expansion [17], watershed transform [18,19], adaptive sweeping [20,21], voronoi division [22], minimal path [23], atlas-based approach [24,25], multilevel B-splines [26] and active contour model [27]. But these lung anatomy-based methods taken a lot of time in pulmonary trachea and vessel segmentation. While the third category typically used deep learning methods to highlight pulmonary fissure representation. Based on this strategy, Gerard et al. presented a cascade FissureNet approach to segment pulmonary fissures [28], but the computational method has a drawback of time consuming in training stage. To overcome the problem, Roy et al. designed a multi-view deep learning network [29] to highlight fissure representation and save time [30]. However, there are no universal methods to remove clutters for pulmonary fissure segmentation [8].

In this paper, a reliable and valuable scheme is presented for pulmonary fissure segmentation. Inspired by previous works [12,13], we utilize the ODoS filter to enhance pulmonary fissures across 2D space. Subsequently, an orientation curvature criterion and an orientation partition scheme are fused to separate candidate fissure profiles from clutters in pre-processing pipeline in orientation field, which is expected to achieve a more complete fissure detection. Moreover, a post-processing pipeline based on shape features is designed by utilizing the difference between tubular structures and pulmonary fissures in magnitude field for pulmonary fissure segmentation.

## 2. Materials and Methods

### 2.1. Data and reference

In this study, 55 CT scans were selected from the Lobe and Lung Analysis 2011 (LOLA11) dataset [31], which were acquired from different scanners and protocols. To evaluate the performance of our scheme, we regarded the fissure references which were verified by two medical experts [12] as the ground truth to evaluate the proposed scheme.

### 2.2. Overview of the proposed scheme

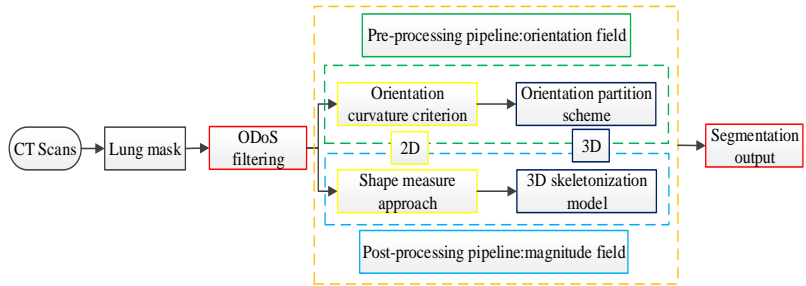

Figure 1: The framework of the computational scheme.

In this paper, we present a reliable and valuable method for fissure segmentation based on ODoS filter and shape features. To reduce the impact of non-lung tissues, lung masks [12,13,32] are used to extract lung regions in advance. The flow chart of our scheme is shown in Figure 1.

## 2.3. ODoS filter

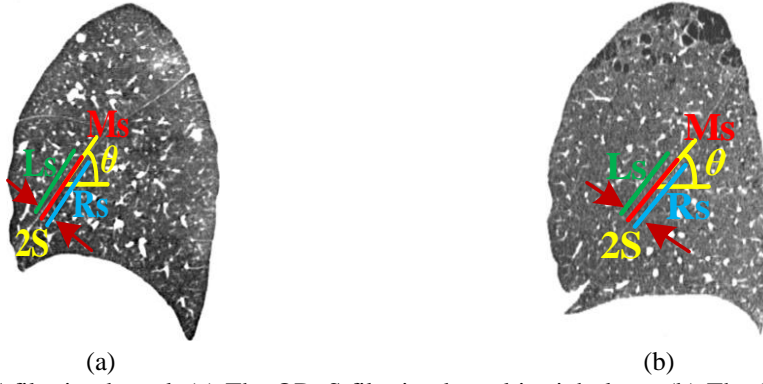

(a) (b)

Figure 2: The ODoS filtering kernel. (a) The ODoS filtering kernel in right lung. (b) The ODoS filtering kernel in left lung.

Due to the poor detection of weak and abnormal fissures, Peng et al. [13] have presented an ODoS filter to highlight fissure representation. The main idea is to take advantage of the intensity and orientation difference between fissures and its surrounding tissues. As shown in Figure 2, the ODoS filtering kernel [12,13] is composed of the left (Ls), middle (Ms) and right (Rs) sticks with different colors, where $\theta$ and $S$ respectively represent the filtering orientation and the inter-stick spacing. To better express the ODoS filter, the mean intensity along the left, middle and right sticks are respectively marked with $u_L$, $u_M$ and $u_R$. The nonlinear differentials perpendicular to the filtering kernel were defined as [12]

$$\lambda_{\perp,\max/\min}^{S,\theta}(x) = \max/\min(u_M - u_L, u_M - u_R) \tag{1}$$

Where $x$ denotes the spatial location in CT images.

To suppress tubular structures, a tubular structure suppression operator was defined [12]

$$\lambda_{\parallel}^{S,\theta}(x) = \sqrt{E(I_j^2) - (E(I_j))^2} \tag{2}$$

Therefore, the fissure line strength measure can be described as

$$l_{\max/\min}^{S,\theta}(x) = \lambda_{\perp,\max/\min}^{S,\theta}(x) - \kappa * \lambda_{\parallel}^{S,\theta}(x) \tag{3}$$

Here, $\kappa$ is equal to 0.7 [12].

Considering the fact that the intensity of fissures is greater than its surrounding tissues, the 2D line strength measure function was defined by Xiao et al. [12]

$$F_{\max/\min}(x) = \max(\max_{1 \leq i \leq 2(L-1)}(l_{\max/\min}^{S,\theta_i}), 0) \tag{4}$$

Where $L=11$ and $\theta$ represent the stick length and orientation, respectively. Subsequently, a cascaded scheme was established to enhance pulmonary fissures and suppress pathological abnormalities. Mathematically

$$F_o(x) = F_{\max}(x) o F_{\min}(x) \tag{5}$$

Here, $o$ indicates the cascading operator [12]. The response $F_o$ comes from the sagittal, axial and coronal cross-sections is marked with $F_S$, $F_A$ and $F_C$, respectively.

Motivated by the reality that $F_{\max}$ plays the major role in $F_o$ for pulmonary fissure enhancement, the orientation response was described as following [13]

$$\theta_{max} = \underset{1 \leq i \leq 2(L-1)}{\arg\max} (l_{max}^{S,\theta_i}) \tag{6}$$

A vector representation

$$\vec{V}_{max}(\theta_{max}) = (\cos\theta_{max}, \sin\theta_{max}) \tag{7}$$

Therefore, the vector $\vec{V}_{max}$ come from sagittal, axial and coronal cross-sections are marked with $\vec{V}_{max}^S$, $\vec{V}_{max}^A$ and $\vec{V}_{max}^C$, respectively.

Inspired by the geometric representation of the vesselness filter [33,34], a shape-tuned response was defined as

$$F^{3D} = (F_o^A + F_o^S + F_o^C) * \frac{median(F_o^A, F_o^S, F_o^C)}{max(F_o^A, F_o^S, F_o^C)} \tag{8}$$

As a result, the intensity and orientation response of the ODoS filter can be fused into a vector form

$$\vec{F}(\theta_{max}) = F^{3D} * \vec{V}_{max}(\theta_{max}) \tag{9}$$

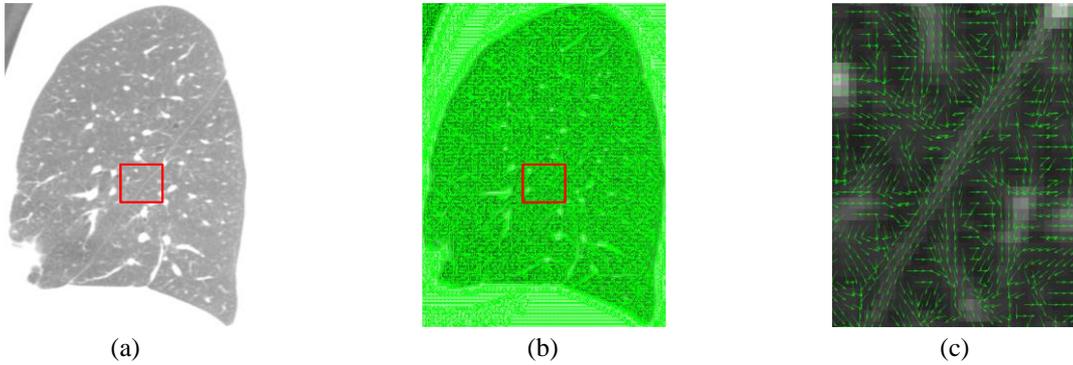

(a)        (b)        (c)

Figure 3. Vector field. (a) The original image. (b) The corresponding vector field. (c) The zoomed rectangle region.

To illustrate the validation of the ODoS filter, a sagittal slice and its corresponding vector field are given in Figure 3(a) and 3(b). Figure 3(c) denotes the zoomed rectangle region of Figure 3(b). As observed, the vector field have been regularized by the ODoS filter. Subsequently, a minimal threshold $T=1$ [13] was selected to avoid some fissures to be eliminated as clutters. Mathematically

$$\vec{F}_v(\theta_{max}) = \begin{cases} \dfrac{\vec{F}(\theta_{max})}{F^{3D}}, & F^{3D} > T \\ 0, & others \end{cases} \tag{10}$$

Different from traditional methods, we adopt an ODoS filter by merging the magnitude information and orientation information to highlight pulmonary fissure representation, which can effectively distinguish between pulmonary fissures and clutters.

### 2.4. Pre-processing pipeline

Although the above operations have a perfect performance in pulmonary fissure enhancement, some undesired structures like vessels and airways still can not be eliminated. To eliminate the undesired structures, an orientation curvature criterion and an orientation partition scheme are fused to separate candidate fissure profiles from clutters in orientation field.

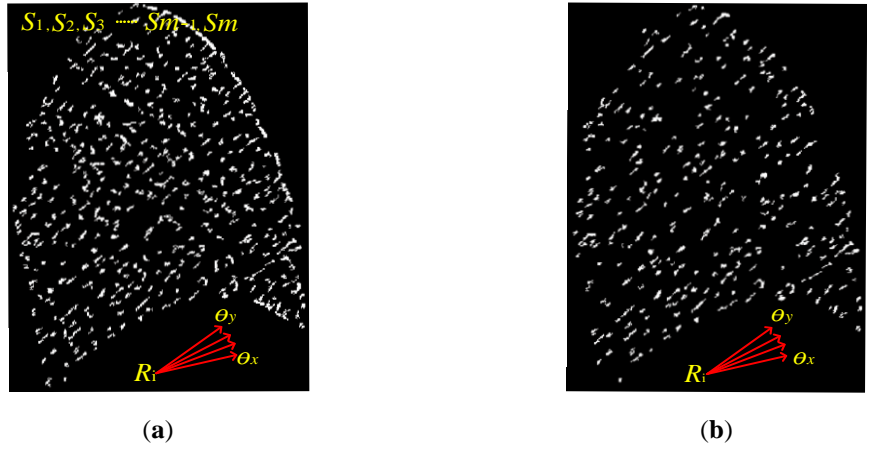

Figure 4. Orientation curvature criterion. (a) Sub-region $R_i$. (b) The filtering results after using the orientation curvature criterion.

In order to suppress clutters, the normalized vector $\vec{F}_v(\theta_{max})$ is divided into $n=8$ [13] overlapped sub-regions and marked with $R_1$, $R_2$, ..., $R_{n-1}$ and $R_n$. Mathematically

$$\vec{F}_{vi} = \begin{cases} \vec{F}_v(\theta_{max}), & \theta_{max} \in R_i \text{ and } |\vec{F}_v(\theta_{max})| > 0 \\ 0, & others \end{cases} \quad (11)$$

In each sub-region $R_i$, pulmonary fissures appear as linear structures in 2D space. Therefore, pulmonary fissures have similar orientations in 2D space. In which, all objects in the sub-region are marked with $S_1$, $S_2$, $S_3$,..., $S_{m-1}$ and $S_m$, the corresponding orientations are labeled with $\theta_1$, $\theta_2$, $\theta_3$,..., $\theta_{m-1}$ and $\theta_m$. We consider the object with the orientation $\theta_i$ is belonging to $[\theta_x, \theta_y]$ as fissures and others as clutters in 2D space. In which, $[\theta_x, \theta_y]$ is the corresponding orientations of the sub-regions $R_i$. Based on this theory, an orientation curvature criterion is defined as

$$\theta_i(S) \subset [\theta_x, \theta_y] \quad (12)$$

After using the orientation curvature criterion, mounts of clutters have been removed. As shown in Figure 4, the approach has a good performance in clutters suppression.

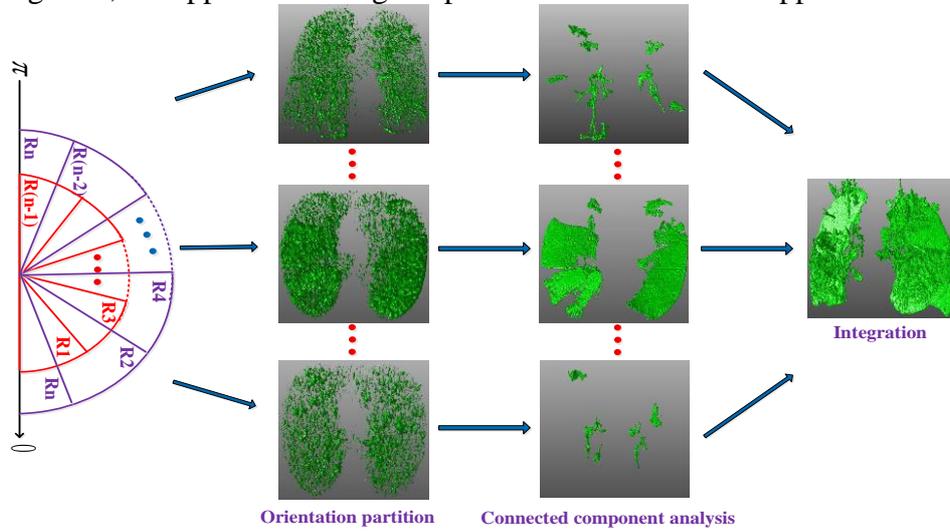

Figure 5. orientation partition scheme

In the same way, in each sub-region $R_i$, pulmonary fissures appear as planar structures in 3D space, an orientation partition scheme is adopt to remove clutters. As shown in Figure 5, in each sub-region, pulmonary fissures appear as planar structures and clutters appear as small structures due to their shape features. Based on this theory, a connected component analysis [35, 36] is used to select candidate fissures and eliminate small structures. Finally, all of the candidate fissures are integrated to form a complete fissure patch.

Unfortunately, Peng et al. [13] used only the sagittal information alone to detect pulmonary fissures, the approach may cause parts of fissures to be undetected. To copy with the problem, we integrate the sagittal, axial and coronal information to make up for the drawback.

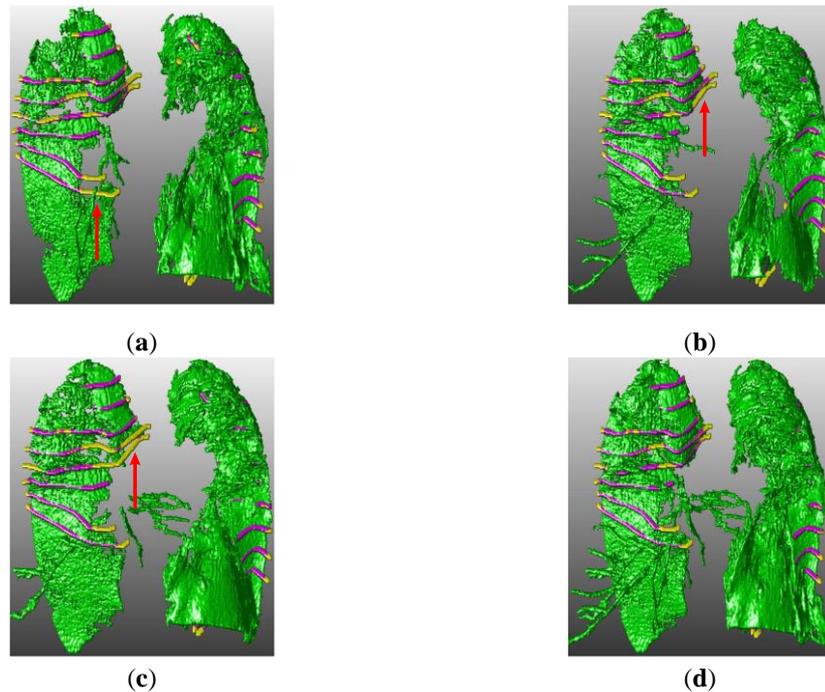

(a) (b)
(c) (d)

Figure 6. Integrating the sagittal, axial and coronal information. (a) Sagittal information. (b) Axial information. (c) Coronal information. (d) Integration.

The detected fissure patches by respectively using the sagittal, axial and coronal information to enhance pulmonary fissures are shown in Figure 6(a), 6(b) and 6(c). Figure 6(d) is the corresponding integrated results. It can be seen that the approach can effectively detect complete fissures. Using only sagittal information may cause parts of fissures to be undetected.

**2.5. Post-processing pipeline**

After using the pre-processing pipeline, some undesired structures like vessels and airways still can't be eliminated. To solve the problem, a shape measure approach and a 3D skeletonization model are combined to segment pulmonary fissures for clutters removal in magnitude field.

As we all know that pulmonary fissures appear as curvilinear profiles in sagittal slice and clutters like airways and vessels appear as tubular structures in 2D space, a matlab function 'regionprops' was used to acquire the selected object property for tuber structure removal. In this section, $K_1$, $K_2$, $K_3$,… $K_{n-1}$ and $K_n$ were used to represent the selected object, $H$ was the major axis length of $K_i$, $W$ was the minor axis length of $K_i$. To remove tuber structures, a shape measure approach [11,37] was used:

$$W(K_i)/H(K_i) \geq T_s \qquad (13)$$

Where $T_s$ is a threshold value. The purpose is to remove tubular structures.

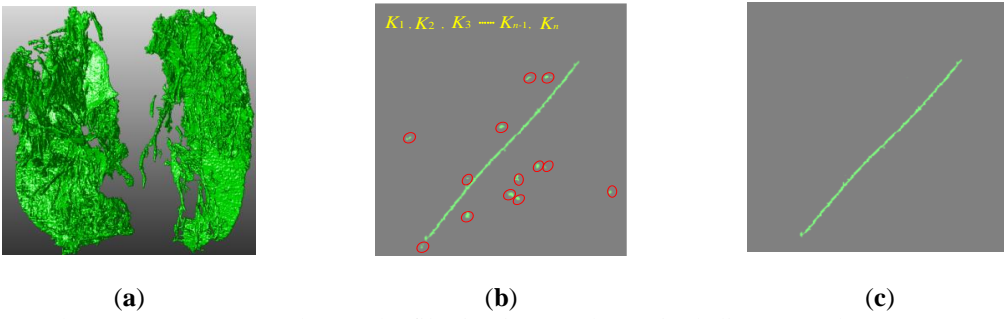

(**a**)            (**b**)            (**c**)

Figure 7. Shape measure approach. (a) The filtering image. (b) Sagittal slice. (c) Tuber structure removal.

As shown in Figure 7, parts of clutters are labeled with red circles in Figure 7(b). After using the shape measure approach, the clutters are removed from the filtering image, and the final results are labeled with $Q$.

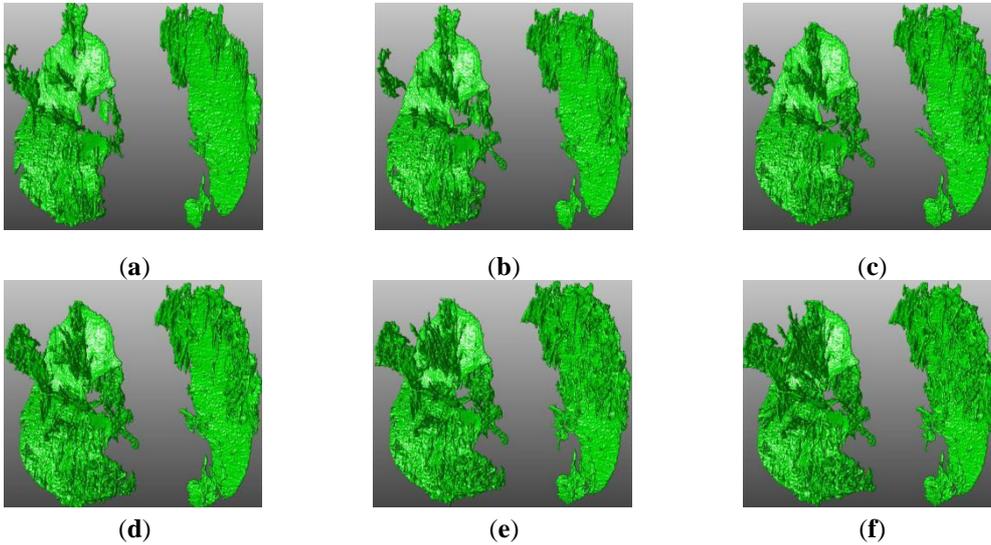

(**a**)            (**b**)            (**c**)

(**d**)            (**e**)            (**f**)

Figure 8. Shape measure approach. (a) $T_s = 0.4$. (b) $T_s = 0.5$. (c) $T_s = 0.6$. (d) $T_s = 0.7$. (e) $T_s = 0.8$. (f) $T_s = 0.9$.

As shown in Figure 8, the threshold $T_s$ is too large, it may result in a lot of clutters to be un-removed. On the contrary, the threshold $T_s$ is too low, it may cause parts of fissures as clutters and removed. To copy with the problem, a clutter removal method based on 3D skeletonization model [38,39,40] is presented to achieve a complete fissure segmentation. The main idea is to select fissure profiles from clutters by breaking their connectivity with the 3D skeletonization model. The detailed algorithms mainly consists of five steps: 3D skeletonization, branch-point removal, connected component analysis, hole-filling and fissure repairing.

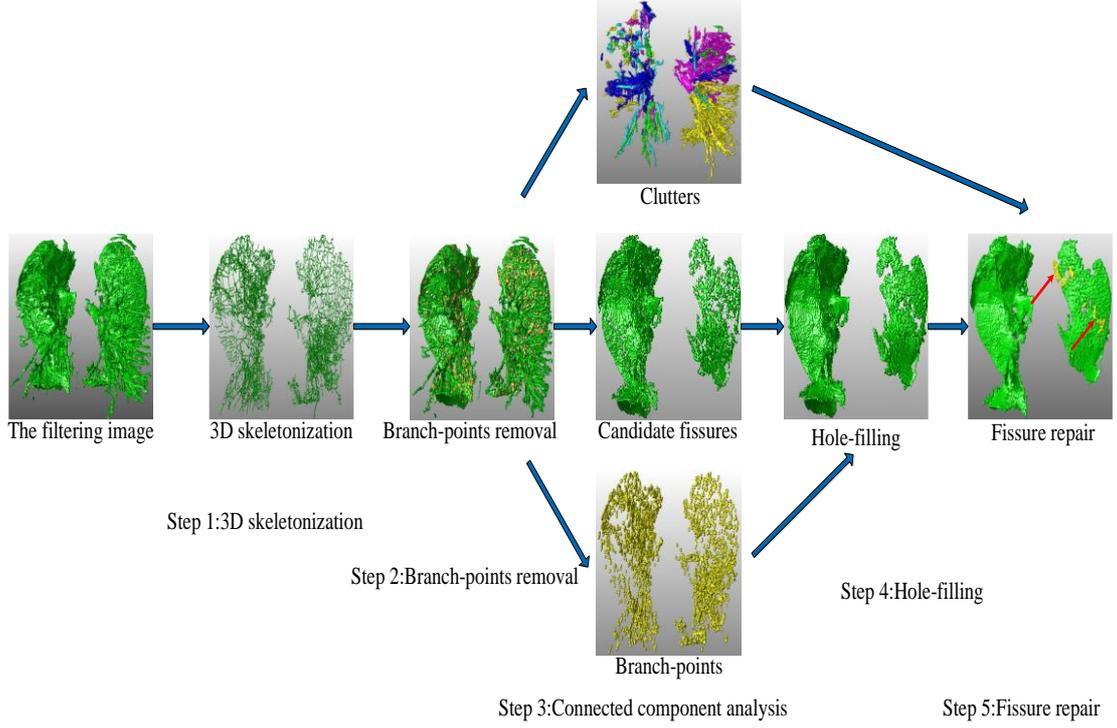

Figure 9. Clutter removal method

Step 1. The 3D skeletonization model is adopted to thin the filtering image $Q$ by computing the complex medial axis of the objects. The purpose is to ensure the invariance of the objects topological structures and geometric features. Therefore, the 3D Euler criterion $X(Q)$ is employed by the global formula

$$X(Q) = O(Q) - H(Q) + C(Q) \tag{14}$$

In which, $Q$ is the filtering image, $O(Q)$, $H(Q)$ and $C(Q)$ are the numbers of connected objects, holes and cavities of $Q$. The filtering image $Q$ can be treated as a finite collection of points. Therefore, a local formula $G(Q)$ can be exploited to reduce the computational complexity of the global formula $X(Q)$. the local formula $G(Q)$ from the algebraic topology can be translated as

$$G(Q) = v - e + f - t \tag{15}$$

Where $v$, $e$, $f$ and $t$ represent the number of vertices, edges, faces and octants in $Q$.

In order to ensure the invariance of the objects topological structures and geometric features for thinning operations, the change of the Euler characteristic $\delta$ is useful in the sense of the Euler criterion in 3*3*3 cube

$$\delta G(Q) = \frac{1}{8} - \frac{\delta e}{4} + \frac{\delta f}{2} - \delta t = 0 \tag{16}$$

Where $\delta t$, $\delta f$ and $\delta e$ represent the changes in the number of octants, faces and edges in 3*3*3 cube. In this section, an Euler table for six-connected objects is adopted to extract the medial axes of the filtered CT images $Q$, the skeletonized image is marked with $Q_k$. As shown in Figure 9, airways and vessels are thinned into single-pixel structures. Generally, the branching regions among fissures, airways and vessels are thinned into branch-points. In other words, the branch-points are removed from the filtered binarized image, pulmonary fissures, airways and vessels are naturally separated from each other.

Step 2. A practical and useful approach is used for branch-points removal in the skeletonized image $Q_k$. In 3D CT images, pulmonary fissures look like planar structures, whereas airways and vessels look like tubular structures. Based on this reality, a simple but effective approach is defined

$$|(N_{26}(p))\cup Q_k|\geq 4 \qquad (17)$$

where $p$ and $N_{26}$ denote the pixel in the skeletonized objects $Q_k$ and its 26 neighborhood regions, respectively. As shown in Figure 10, Figure 10(a) and (c) represent the tubular structures, Figure 10(b) and (d) represent the bifurcated structures, the green dots and yellow dots respectively denote $p$ and its 26 neighborhood regions. If there are four or more points within a 3*3*3 cube ($p$ and its 26 neighborhood), we regard the pixel $p$ as the branch-point. After removing all of branch-points, the complex branching structures like airways and vessels are naturally divided into a series of small fragments.

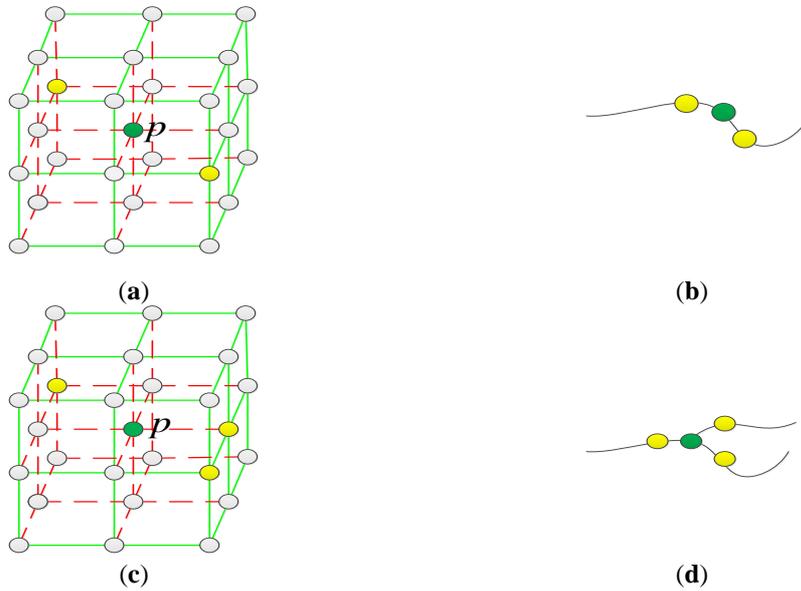

Figure 10. Branch-points removal.. (a) and (b) No branch-points. (c) and (d) Branch-points.

Step 3. The connected component analysis is used to select the candidate fissures. After removing the branch-points, pulmonary fissures still remain good connectivity, whereas airways and vessels are divided into small fragments. Based on this strategy, the connected component analysis is adopted to remove clutters. Generally, the branching regions among fissures, airways and vessels are thinned into branch-points. In other words, the branch-points are removed from the filtered image, pulmonary fissures, airways and vessels are naturally separated from each other.

Step 4. A hole-filling algorithm is used to achieve a complete segmentation. There are a large number of holes in the candidate fissure surfaces. To circumvent the problem, we put the previously eliminated branch-points back to fill the holes. At last, a volume sorting scheme [12,13] is employed to isolate fissure profiles for pulmonary fissure segmentation. The segmented fissure is marked with $Q_S$.

$$Q_L = Q - Q_S \qquad (18)$$

As a result, clutters are treated as bigger objects in the rest objects $Q_L$, the bigger objects $B_1 = \max(Q_L)$, $B_2 = \max(Q_L - B_1)$, $\cdots$, $B_n$. Therefore, small objects in the rest image $Q_L$ are preserved

$$Q'_L = Q_L - \sum_{i=1}^{n} B_i \tag{19}$$

Subsequently, the small objects $Q'_L$ and the segmented fissure $Q_S$ are integrated into many objects, mathematically

$$Q''_L = Q'_L + Q_S \tag{20}$$

Finally, a connected component analysis approach is used to select bigger objects from $Q''_L$ as the final segmented pulmonary fissures. As shown in Figure 9, the missed fissure (labeled with red arrows) can be repaired.

## 3. Experimental results

### 3.1. Evaluation criteria

In this study, A 3mm width around the segmented fissure is labeled with $S_1$, whereas a 3mm width around the corresponding ground truth is marked with $R_1$. The overlapped areas between the segmented fissure and $R_1$ are treated as $TP_1$, and the rest being $FP$. In a similar way, the overlapped areas between the ground truth and $S_1$ are treated as $TP_2$, and the rest being $FN$. Accordingly, the False Discovery Rate ($FDR$), False Negative Rate ($FNR$) and $F_1$-score ($F_1$) are defined as

$$FDR = FP/(TP_1 + FP) \tag{21}$$

$$FNR = FN/(TP_2 + FN) \tag{22}$$

$$F_1 = \frac{2(1-FDR)(1-FNR)}{2-FDR-FNR} \tag{23}$$

### 3.2. Visual inspection

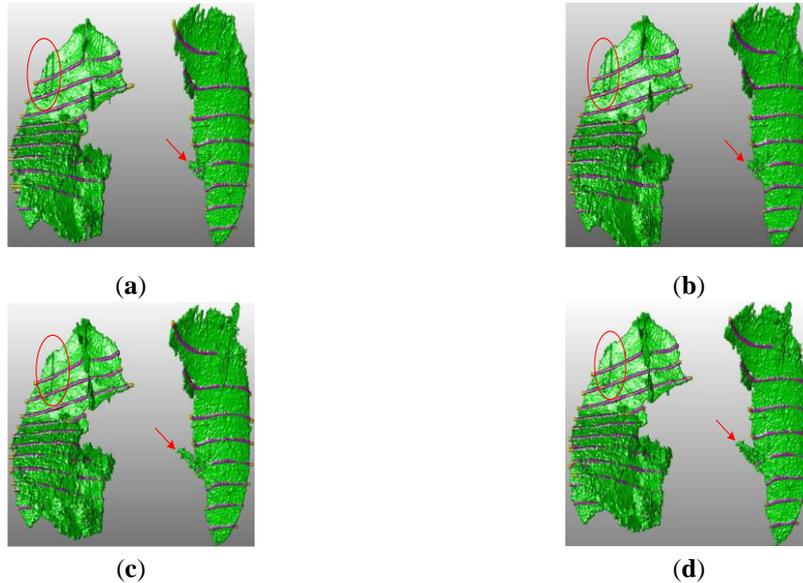

Figure 11. Different threshold $T_s$. (a) $T_s = 0.4$. (b) $T_s = 0.5$. (c) $T_s = 0.6$. (d) $T_s = 0.7$.

To select an optimal threshold $T_s$ in shape measure approach, one representative segmentation is chosen with different threshold. As shown in Figure 11, the threshold is too low, it may cause parts fissures (marked with red ellipses) to be undetected. On the contrary, the threshold is too large, it may cause parts clutters (marked with red arrows) to be un-removed.

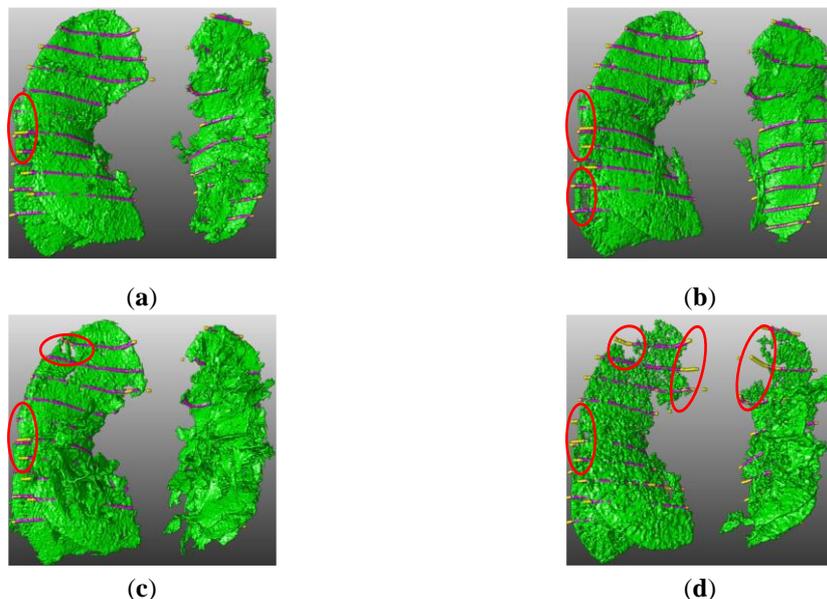

Figure 12. Pulmonary fissure segmentation with different method. (a) The proposed method. (b) ODoS filter [13]. (c) DoS filter [12]. (d) Fissureness filter [26].

To demonstrate the segmentation performance, we compared the computerized scheme against three typical methods [12,13,26]. As shown in Figure 12, Figure 12(a), (b), (c) and (d) respectively give the computerized scheme, ODoS [13], DoS [12] and fissureness [26] filtering segmentation. The segmentation results, the ground truth and their overlapped areas are rendered in green, yellow and purple. The benefits of the computerized scheme can be discovered in the areas marked with red ellipses, the weak and abnormal fissures can be found by our scheme and lead to a lower FNR value. Experimental results show that the computerized scheme performs well in weak and abnormal fissure segmentation.

### 3.3. Quantitative evaluation

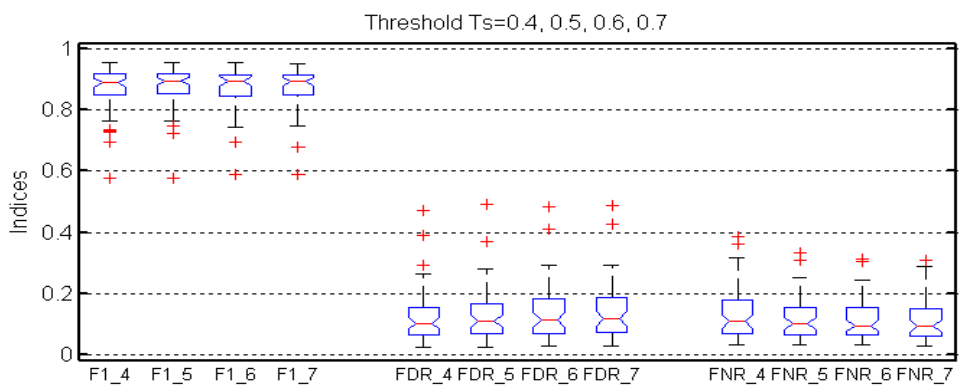

Figure 13. Quantitative evaluation of the computational scheme with different threshold $T_s$.

The box-plots of indices corresponding to the computational scheme with different threshold $T_s = 0.4$, $T_s = 0.5$, $T_s = 0.6$ and $T_s = 0.7$ are shown in Figure 13. The median values are 0.891, 0.896, 0.896, 0.894, 0.102, 0.109, 0.111, 0.116, 0.107, 0.100, 0.093, and 0.092. Both visual inspection and quantitative evaluation illustrated that the computational method has a good performance in pulmonary fissure segmentation.

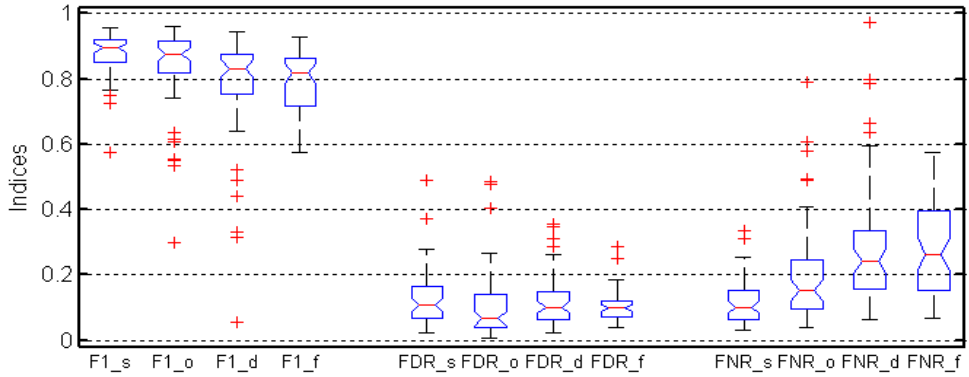
Figure 14. Segmentation validation on LOLA11 dataset with different methods.

In Figure 14, The box-plots of indices corresponding to the computational scheme (s), ODoS (o), DoS (D), and fissureness (f) filtering scheme are drawn next to each other. The median values are 0.896, 0.877, 0.833, 0.818, 0.109, 0.065, 0.102, 0.098, 0.100, 0.152, 0.242, and 0.264. Both visual inspection and quantitative evaluation illustrated that the computational method can outperform the compared methods [12,13,26].

Table 1. The indices of different methods in publicly LOLA11 dataset.

| Methods | $F_1$ | FDR | FNR |
| --- | --- | --- | --- |
| Proposed | **0.896** | 0.109 | **0.100** |
| Method[10] | 0.582 | 0.087 | 0.566 |
| Method[11] | 0.892 | 0.085 | 0.117 |
| Method[12] | 0.833 | 0.102 | 0.242 |
| Method[13] | 0.877 | 0.065 | 0.152 |
| Method[14] | 0.687 | 0.087 | 0.451 |
| Method[26] | 0.818 | 0.098 | 0.264 |
| Method[30] | 0.890 | No | No |

In Table 1, compared with different methods, the proposed method has the largest median F1 values, which indicates that the presented method can efficiently segment fissures in 3D CT images. The proposed method has the lowest median FNR value, Which means that the presented method has a good performance in detection of weak and abnormal fissures.

## 4. Discussion

In this study, a computerized scheme based on ODoS filter and shape features is introduced for pulmonary fissure segmentation. In methodology, our scheme has many unique merits and characteristics. First, an ODoS filter based on vector field instead of the intensity field is developed to enhance pulmonary fissures, which can accurately discriminate between fissure profiles and other tissues. Second, an orientation curvature criterion and an orientation partition scheme are modified to highlight fissure representation and suppress clutters. While the original orientation partition scheme may cause parts of weak and abnormal fissures to be undetected. Third, the post-processing pipeline is performed on pulmonary fissure segmentation under a 3D branch-point removal framework. Unlike traditional methods [9, 12] that worked in 2D space, this operation makes clutters reasonably expelled. In addition, under the help of the improved ODoS filter and 3D skeletonization model, our scheme are expected to preserve the integrity of weak and abnormal fissure segmentation.

The computerized scheme outperformed with these typical methods [10, 11, 12, 13, 14, 26, 30]. Experimental results show that our scheme performs well in weak and abnormal fissure segmentation. Compared with ground truth, the proposed method obtained a higher F1-score of 0.896 than the compared methods. The reason is that the computerized scheme used the sagittal, diagonal and coronal information to detect weak and abnormal fissures, then a 3D branch-point removal framework was designed to segment fissures. While Peng et al. [13] employed only the sagittal information, it may cause some fissures to be undetected. On the Contrary, Xiao et al. applied only the intensity information to segment fissures, parts of fissures were regarded as clutters and removed [12]. Using a different way, Doel et al. utilized a 3D vessel distance transform to suppress clutters [26]. Unfortunately, parts of vessels cross fissures, it may cause some fissures simultaneously suppressed.

However, the computerized scheme has many shortcomings and disadvantages. The fatal limitation of our scheme is a longer time computation. Besides, parts of pathological clutters appear as planar structures, which cannot be eliminated by the computerized scheme. Although our scheme has many drawbacks, the integrity of the fissure segmentation have been greatly improved under the fissure segmentation framework. In the future, many methods like deep learning framework[41, 42], prior knowledge-based segmentation [43] and feature extraction[44] may be adopted to improve the presented framework. Especially, the pathological clutters removal will be the focus of our attention.

## 5. Conclusion

In this paper, a reliable and valuable computerized scheme is presented for fissure segmentation. Considering the reality that pulmonary fissures look like line-curved structure in 2D space and planar structure in 3D space, an ODoS filter and an orientation partition scheme are developed to highlight pulmonary fissures and suppress clutters, which are expected to achieve a more complete fissure detection. Another contribution of our scheme focus on the processing pipeline. Using an ingeniously designed processing pipeline to isolate weak and abnormal fissure patches. Compared with seven typical methods, our scheme can achieve a higher segmentation accuracy.

**Data Availability**
The LOLA11 dataset can be download in the website https://lola11.grand-challenge.org/.

**Conflict of Interest**
The authors declare no conflicts of interest.

**Authors' Contributions**
Conceptualization, Yuanyuan Peng and Hongbin Tu; methodology, Yuanyuan Peng and Pengpeng Luan; validation, Yuanyuan Peng and Xiong Li; formal analysis, Xiong Li; data curation, Yuanyuan Peng and Xiong Li; writing original draft preparation, Yuanyuan Peng; funding acquisition, Yuanyuan Peng, Hongbin Tu and Xiong Li; paper modification, Yuanyuan Peng and Ping Zhou. All authors have read and agreed to the published version of the manuscript.

**Acknowledgments**
This research was supported by the Jiangxi Provincial Natural Science Foundation (nos. 20212BAB202007, 20202BAB212004, 20212BAB211009, 20204BCJL23035, 20192ACB21004, 20181BAB202017), the Hunan Provincial Natural Science Foundation (no. 2021JJ30165), the Hunan Special Funds for the Construction of Innovative Province(Huxiang